\documentclass[trackchanges,twocolumn,onecolappendix,astrosymb]{aastex701}

\usepackage{amsmath}
\DeclareMathOperator{\sinc}{sinc}
\usepackage{tikz}
\usetikzlibrary{arrows.meta,calc,angles,quotes}
\usepackage{adjustbox}

\begin{document}

\title{On the angular localization of gravitational-wave signals by pulsar timing arrays}

\author[0000-0001-8217-1599]{Stephen R. Taylor}
\affiliation{Department of Physics \& Astronomy, Vanderbilt University,
2301 Vanderbilt Place, Nashville, TN 37235, USA}
\email{stephen.r.taylor@vanderbilt.edu}

\begin{abstract}
We provide a complete study of the factors influencing gravitational-wave signal localization using pulsar timing arrays. We derive analytical expressions for the Cram\'er-Rao sky localization precision that delineate the impact of the angular proximity, $\xi$, between the pulsar and the gravitational wave source, and the precision, $\sigma_L$, with which pulsar distances are known. Interference between the Earth and pulsar terms creates rapid angular oscillations for sky-coordinate Fisher matrix elements that aid localization, which is complemented by more broadly varying antenna response gradient information. The relative importance of these factors depends on whether pulsar distances are known precisely [i.e., $\sigma_L\leq\lambda_\mathrm{GW}/(1-\cos\xi)$] or imprecisely. If pulsar distances are known precisely, tightening this distance precision improves signal localization according to $\Delta\Omega_\mathrm{sky}\propto(\sigma_L/L)^2$ until the Earth-pulsar system reaches its diffraction limit, $\Delta\Omega_\mathrm{sky}\propto(\lambda_\mathrm{GW}/L)^2$. If pulsar distances are not known well, localization precision is degraded, but more pulsars in close proximity to the source is the best means of improving. With $\alpha$ indexing pulsars, this scales as $\Delta\Omega_\mathrm{sky}~\propto~(\sum_\alpha \mathrm{SNR}_\alpha^2/\xi_\alpha^2)^{-1}$ in the small-angle limit of the unmarginalized Fisher matrix, and we derive the analytic generalization to any angle between a pulsar and the source. Finally, we study a scenario where pulsar-term phases are treated as nuisance variables that are unconnected to binary or PTA properties. This phase-decoupled scenario, which is how all PTA continuous wave searches are currently conducted, delivers localization performance similar to the antenna-response--driven case, and does not exhibit significant improvement as pulsar distance precisions are tightened.
\end{abstract}



\section{Introduction}

Precision timing of Galactic millisecond-period pulsars enables dedicated long-term campaigns to detect the influence of nanohertz-frequency gravitational waves (GWs). When GWs propagate through the Galaxy and pass through the spacetime between a pulsar and the Earth, they induce a distortion that causes pulse arrival times to deviate from modeled expectations \citep{1978SvA....22...36S,1979ApJ...234.1100D,1990ApJ...361..300F}. By modeling the relationship between these timing residuals across all pulsars in an observed ensemble, pulsar timing array (PTA) collaborations search for stochastic and deterministic sources of GWs. 

These efforts have recently seen significant success in delivering strong evidence for a background of nanohertz-frequency GWs \citep{NG15_gwb,2023A&A...678A..50E,2023ApJ...951L...6R,2023RAA....23g5024X,MPTA_gwb}, likely from a large population of tightly-orbiting supermassive black-hole binary (SMBHB) systems. This breakthrough has encouraged even greater attention on the field's next milestone: fragmenting this signal into a number of extremely massive or nearby binaries that are individually resolvable, contrasted against a Gaussian signal from the remaining unresolved ensemble.

The search for GW signals from individual binary systems with PTAs is arguably more challenging than for a stochastic GW background. The latter requires searching for an interpulsar-correlated red stochastic signal amid intrinsic low-frequency pulsar noise and time-variable interstellar-medium processes. Yet the signal model for an individual SMBHB requires one to account for the dynamical state of the binary when the GW first passes each pulsar, then passes Earth. The latter signal component has a common phase across all pulsars, but the former depends on the geometry between the Earth, pulsar, and GW source. In the simplest scenario, this expands the search dimensionality from eight binary parameters (sky location, orbital frequency, chirp mass, GW polarization, initial phase, distance, orbital inclination) to also include the distance to each pulsar.

\citet{2011MNRAS.414...50D} used a proof-of-principle study with precisely-measured pulsar distances to demonstrate how a binary source in 3C66B could be localized to within $\sim10^{-5}$~deg$^2$, albeit leveraging GW wavefront curvature effects. Their model was explicitly \textit{phase-linked} across the array, with pulsar-term phases depending on pulsar distances, as well as PTA and GW source characteristics. The modern Bayesian approach to these continuous GW searches began with \citet{2010arXiv1008.1782C}, who included the pulsar term in their phase-linked signal model, with pulsar distances allowed to be inferrable parameters (subject to prior constraints). They noted the central challenge as one of hopping around numerous secondary maxima in pulsar distances that all appear to blend together\footnote{See also the following presentation by N. Cornish from the conference \textit{Astro-GR@Mallorca 2011}:  \href{https://drive.google.com/open?id=0B5DjnMovgoFeRGZfbTIxa0dDRlU}{What can we learn about Black Holes using Pulsar Timing Arrays?}}. However, GW source localization precision was shown to be approximately an order of magnitude better than Earth-term--only studies \citep{2010PhRvD..81j4008S}. Later, \citet{neil_2011_presentation} and \citet{2013CQGra..30v4004E} independently developed efficient mechanisms of navigating this complicated likelihood surface of pulsar distances and GW source location through a sequence of big and small distance jumps.

In a subsequent variation, formalized in \citet{2014PhRvD..90j4028T}, one may abandon any attempt to phase connect the pulsar terms across the array. In this \textit{phase-decoupled} approach, the pulsar term phases are merely nuisance parameters over which we marginalize, leaving the pulsar distance to only influence the frequency and amplitude of the pulsar term. While it expands the search dimensionality to eight plus twice the number of pulsars, this is how production-level PTA searches are currently conducted. 

These Bayesian pipelines have been continually refined to accommodate increasing numbers of pulsars and observations using a variety of advanced sampling schemes and bespoke proposal mechanisms \citep{2020CQGra..37m5011B,2022PhRvD.105l2003B,2024CQGra..41v5017B,2025PhRvD.112h3035G}. Complementing the Bayesian approach are several frequentist statistics that are constructed through analytic maximizations over groupings of binary parameters, \citep{2012PhRvD..85d4034B,2012ApJ...756..175E,2015MNRAS.449.1650Z} or null-stream frameworks \citep{2016arXiv160703459H,2018MNRAS.477.5447G}.

The importance of accounting for pulsar terms in PTA GW searches is well studied. Given the limited binary evolution likely over typical PTA observation timelines, the fact that a lagged imprint of a GW source's past dynamical state is contained within the structure of the induced timing residuals allows for the binary's chirping behavior to be constrained and its chirp mass to be measured. Moreover, neglecting pulsar terms in all-sky searches can lead to substantial localization bias \citep{2016MNRAS.461.1317Z}. 

But if they can be fully leveraged, pulsar terms are the key to true precision GW source localization with PTAs. \citet{2010PhRvD..81j4008S} used numerical Fisher estimates with an Earth-term signal model to study PTA parameter estimation of single GW sources, observing the importance of pulsar proximity to localization when only antenna response patterns are used. The localization improvements of pulsar-term modeling demonstrated in \citet{2010arXiv1008.1782C} were examined in \citet{2011MNRAS.414.3251L}, who showed that interference between the Earth term and pulsar term induces rapidly oscillating responses on the sky, drastically improving localization beyond the broad capabilities of antenna responses. The analytical study of \citet{2012PhRvD..86l4028B} brought all these factors together within a Fisher framework, explaining some of the scaling behaviors with pulsar proximity when pulsar distances are either well known (interference effects dominate) or poorly constrained (antenna responses dominate). The present paper follows the lineage of the latter, with the aim of providing deeper understanding of the different signal localization scenarios, context to numerical localization results \citep[e.g.,][]{2019MNRAS.485..248G,2024ApJ...976..129P,2023PhRvD.108l3535K,2025arXiv251001317S,2025arXiv251001316P,2026PhRvD.113b2001K,2026A&A...706A.299G}, and more general scaling expressions.

This paper is laid out as follows. We review the GW signal model for a continuous wave in PTAs in \S\ref{sec:sec_signal}. In \S\ref{sec:sec_localization} we outline the Fisher formalism in which we study GW signal localization, and provide analytic derivations of the partial derivatives of the signal model with respect to GW source sky coordinates. We discuss how the sky uncertainty area is computed in terms of the Fisher information matrix, how nuisance parameters are marginalized over, and how different factors influence localization in the scenarios where pulsar distances are known precisely or imprecisely. We assess our analytic results with numerical studies in \S\ref{sec:sec_numerical}, exploring the impact of pulsar proximity to the GW source, pulsar distance uncertainty, source binary chirp mass, and highlighting the role of pulsar terms in the current \textit{phase-decoupled} signal approach. \S\ref{sec:sec_conclusions} has our concluding remarks. In the following, we assume $G=c=1$.

\section{Signal model} \label{sec:sec_signal}

\begin{figure} 
    \includegraphics[width=0.9\columnwidth]{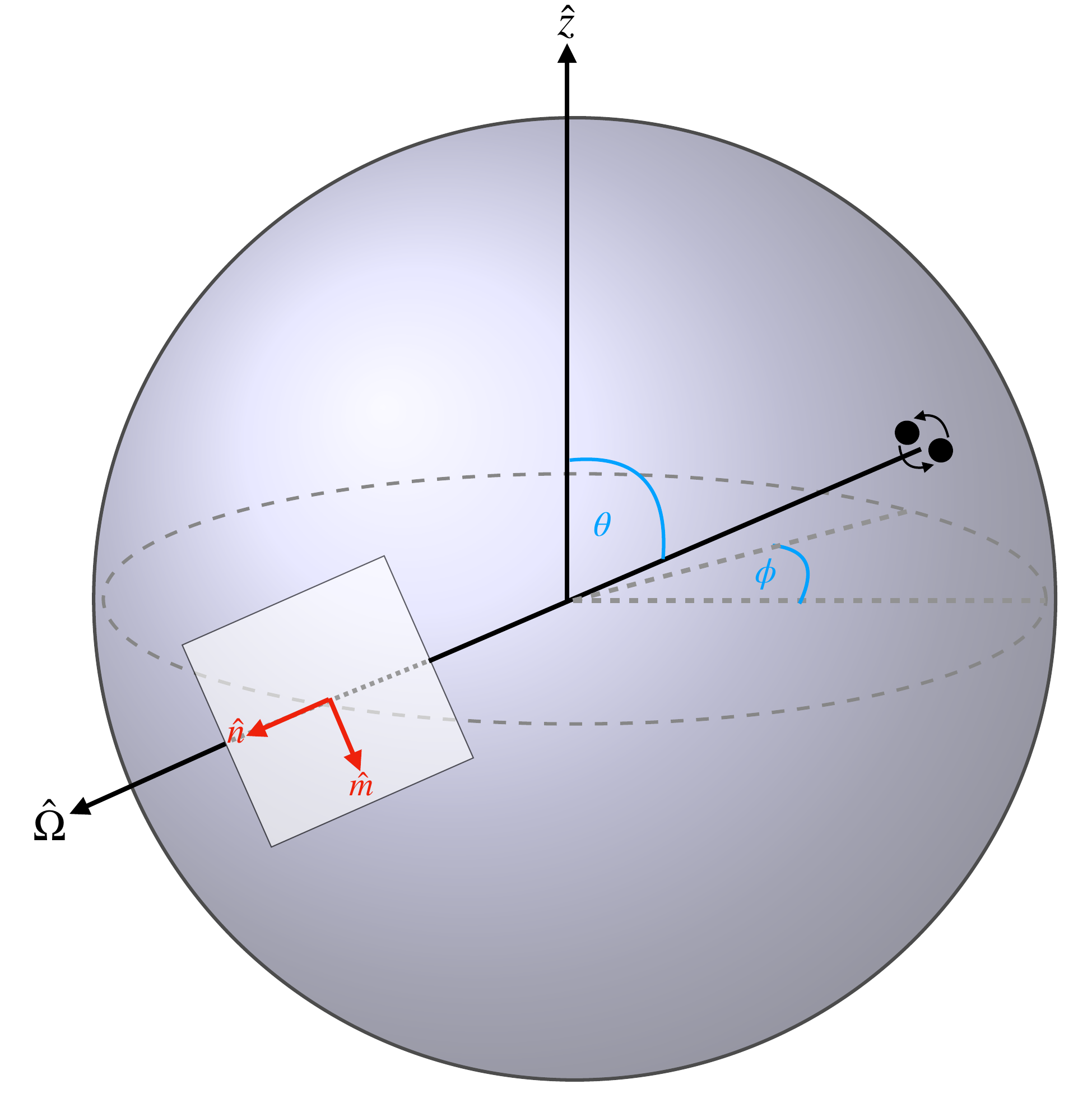}
    \caption{\label{fig:gw_geometry}As is convention, the direction of GW propagation is $\hat\Omega$ such that the unit vector pointing to the source at spherical polar coordinates $(\theta,\phi)$ is $-\hat\Omega$. We define a right-handed orthogonal basis triad such that $\hat\Omega=\hat{m}\times\hat{n}$, where $\hat{m}$ and $\hat{n}$ lie in the plane transverse to the direction of GW propagation.}
\end{figure}

In all that follows, we assume that sources of GWs are far enough away from the Milky Way that we may adopt a plane wave description. A continuous GW signal from a circular SMBHB is modeled in terms of the perturbations it induces to the arrival times of pulses from a given pulsar. In a single pulsar this can be written as,
\begin{equation}
    r(t) = \tilde{F}^+(\hat\Omega,\psi)\Delta r_+(t) + \tilde{F}^\times(\hat\Omega,\psi)\Delta r_\times(t),
\end{equation}
where $\tilde{F}^{\{+,\times\}}$ are GW antenna response functions for the $+$ and $\times$ GW polarizations, which depend on the sky location of the pulsar, the direction of GW propagation $\hat\Omega$, and (in this notation) the GW polarization angle $\psi$. The term $\Delta r_{\{+,\times\}}(t)\equiv r_{\{+,\times\}}(t_p) - r_{\{+,\times\}}(t)$ is the difference between the pulsar term and the Earth term, with $t_p = t - \tau = t - L(1+\hat\Omega\cdot\hat{p})$ where $L$ is the distance to the pulsar and $\hat{p}$ is a unit vector pointing to the pulsar from Earth (or the Solar System barycenter). 

We adapt some of the notation from \citet{2024PhRvD.109h3038A} to write a more compact representation of the residual model in the complex plane, where the measured residuals are the real part of the following expression, $r(t)=\mathbb{R}[s(t)]$:
\begin{equation} \label{eq:evol}
    s(t) = \frac{\tilde{F}}{4i}\left[\, U\,w(t)
                                    -V\,w^*(t)\, \right],
\end{equation}
where,
\begin{equation}
    w = \left( \frac{e^{2i\Phi}}{\omega^{1/3}} - \frac{e^{2i\Phi_{p}}}{\omega_{p}^{1/3}} \right) \equiv w_e - w_p
\end{equation}
such that $\Phi(t)=\omega (t-t_0) + \Phi_0$, $\omega=2\pi f_\mathrm{orb}=\pi f_\mathrm{GW}$ is the Earth-term orbital angular frequency (assumed to be approximately non-evolving over PTA observation spans); and $\Phi_0$ is a reference orbital phase measured at $t_0$. The variables $\Phi_p$ and $\omega_p$ are the orbital phases and the orbital angular frequency, respectively, of the pulsar term. These are similarly related as their Earth-term counterparts, and depend on the pulsar term's lag behind the Earth term, i.e., $\Phi_p(t)=\omega_p (t-t_0) + \Phi_{p,0}$, and
\begin{align}
    \omega_p &= \omega \left[ 1 + \frac{256}{5} \mathcal{M}^{5/3} \omega^{8/3} L(1+\hat\Omega\cdot\hat{p}) \right]^{-3/8}, \nonumber\\
    \Phi_{p,0} &= \Phi_{0} + \frac{1}{32 \mathcal{M}^{5/3}} \left( \omega^{-5/3} - \omega_{p}^{-5/3} \right).
\end{align}

The amplitude factors $U$ and $V$ are given by
\begin{equation}
    U = A(1-\cos\iota)^2, \quad V = A(1 + \cos\iota)^2,
\end{equation}
where $\iota$ is the orbital inclination angle measured between the binary angular momentum vector and a line-of-sight vector to the source, and $A=2\mathcal{M}^{5/3}/ D_L$, such that $\mathcal{M}$ is the (redshifted) binary chirp mass and $D_L$ is the binary luminosity distance.

The usual (real) antenna response functions are collected together as $\tilde{F} \equiv Fe^{-2i\psi} = (F^+ + iF^\times)e^{-2i\psi}$, where $\tilde{F}^+$ and $\tilde{F}^\times$ are the real and imaginary components of $\tilde{F}$, respectively. In this notation, $F$ is given by
\begin{equation} \label{eq:complex_F}
    F = \frac{1}{2}\frac{[(\hat{m}+i\hat{n})\cdot\hat{p}]^2}{(1+\hat\Omega\cdot\hat{p})},
\end{equation}
where $\hat\Omega$ is the direction of GW propagation as before, and $\hat{m}$ and $\hat{n}$ lie in the plane perpendicular to this direction, forming a right-handed orthogonal basis triad such that $\hat\Omega=\hat{m}\times\hat{n}$. If $(\theta,\phi)$ denote the GW source location in spherical polar coordinates, these vectors take the following explicit form in terms of Cartesian unit vectors:
\begin{align}
    \bm{\hat{\Omega}} &= -(\sin\theta \cos\phi) \, \bm{\hat{x}} - (\sin\theta \sin\phi) \, \bm{\hat{y}} - \cos\theta \, \bm{\hat{z}}, \\
    \bm{\hat{m}} &= \sin\phi \, \bm{\hat{x}} - \cos\phi \, \bm{\hat{y}}, \\
    \bm{\hat{n}} &= -(\cos\theta \cos\phi) \, \bm{\hat{x}} - (\cos\theta \sin\phi) \, \bm{\hat{y}} + \sin\theta \, \bm{\hat{z}}.
\end{align}

This geometry is shown in \autoref{fig:gw_geometry}, with the GW origin represented by a cartoon binary (not to scale).

\subsection{Non-evolving binary}

It will be useful for later analytic approximations to consider the case in which a binary's orbital frequency does not evolve significantly over the lag time, $\tau=L(1+\hat\Omega\cdot\hat{p})$, between the pulsar term and Earth term.\footnote{However, all subsequent numerical studies are carried out with full evolution of the pulsar-term frequency.} In this case, $\omega_p\approx\omega$, $\Phi_p \approx \Phi - \omega\tau$, and the residual model can be written more succinctly as 
\begin{equation} \label{eq:nevol}
    s(t) = \frac{\tilde{F}}{4i\omega^{1/3}}\left[ U g(\tau)e^{2i\Phi} - V g^*(\tau)e^{-2i\Phi} \right],
\end{equation}
where $g(\tau) = 1 - e^{-2i\omega\tau}$. In this notation, $g(\tau)$ encodes the interference between Earth and pulsar terms, which is seen more explicitly by rewriting it as
\begin{equation} \label{eq:g_factor}
    g(\tau) = 2i e^{-i\omega\tau} \sin\left[ \omega L(1+\hat\Omega\cdot\hat{p})\right].
\end{equation}

\subsection{Optimal signal-to-noise ratio} \label{sec:snr}

The optimal GW signal-to-noise ratio (SNR) in one pulsar can be defined through a noise-weighted inner product of the signal with itself. In order to keep using the complex residual formalism, we assume that the GW is observed for many cycles, such that $T\gg1/\omega$, and
\begin{equation} \label{eq:snr}
    \mathrm{SNR}^2 = (\mathbb{R}[s]\,|\,\mathbb{R}[s])\approx \frac{1}{2}(s | s) = \frac{1}{2} \vec{s}(t)^\dagger C_n^{-1} \vec{s}(t), \\
\end{equation}
which we take to be additive in quadrature over pulsars in the array, $\mathrm{SNR}^2 = \sum_\alpha\mathrm{SNR}^2_\alpha$. \autoref{eq:snr} is general, but in the analytic steps that follow within this subsection we employ the non-evolving signal model of \autoref{eq:nevol}. Moreover, without loss of generality in our subsequent analytic or numerical studies, we ignore the details of fitting the deterministic pulsar timing model, as well as intrinsic pulsar white, red, and chromatic noise. Hence, we model the noise covariance with a simple white noise term, as $C_{n,\alpha} = \sigma_\alpha^2 \mathbb{I}$.

We now derive the time-integrated, polarization-averaged, and inclination-averaged SNR in a single pulsar. For simplicity, we suppress nested averaging notation and express the result as $\overline{\mathrm{SNR}}$. We begin with
\begin{align}
    \mathrm{SNR}^2_\alpha =& \frac{1}{2}(s_\alpha|s_\alpha) \nonumber\\
    =& \frac{|F_\alpha|^2}{32} \frac{T\omega^{-2/3}}{\sigma^2_\alpha \Delta t} \left[ (Uw_\alpha^* - Vw_\alpha)(Uw_\alpha - Vw_\alpha^*) \right] \nonumber\\
    =& \frac{|F_\alpha|^2}{32} \frac{T\omega^{-2/3}}{\sigma^2_\alpha \Delta t} \left[ (U^2 + V^2)|g_\alpha|^2 \right. \nonumber\\
    &\left.- UV(g_\alpha^*)^2 e^{-4i\Phi} - UV(g_\alpha^2) e^{4i\Phi} \right]
\end{align}
where the GW polarization term cancels through conjugation and can be trivially averaged over, and the exponential terms average out to zero when $T\gg1/\omega$, such that the time-integrated SNR is
\begin{equation}
    \mathrm{SNR}^2_\alpha = \frac{ T \omega^{-2/3}}{32\sigma_\alpha^2\Delta t} ( U^2 + V^2 ) |F_\alpha|^2 |g_\alpha|^2
\end{equation}
where $\Delta t$ is the cadence of observations on pulsar $\alpha$, often fortnightly or monthly. We then note that
\begin{align}
U^2+V^2 &= A^2[(1-\cos\iota)^4 + (1+\cos\iota)^4] \nonumber\\
&= 2A^2[1+6\cos^2\!\iota + \cos^4\!\iota],
\end{align}
such that averaging over inclination angles with $\cos\iota\in[-1,1]$, delivers the final expression
\begin{align}
    \overline{\mathrm{SNR}^2_\alpha} &= 
    \frac{A^2 T \omega^{-2/3}}{5\sigma_\alpha^2\Delta t} |F_\alpha|^2 |g_\alpha|^2 \nonumber\\
    &= \frac{16}{5}\frac{\mathcal{M}^{10/3}}{D_L^2 \omega^{2/3}}\frac{T}{\sigma_\alpha^2\Delta t} |F_\alpha|^2 \sin^2\left[ \omega L_\alpha (1 + \hat\Omega\cdot\hat{p}_\alpha)\right],
\end{align}
where in the final line we have used \autoref{eq:g_factor}. We can further reduce this expression by evaluating $|F_\alpha|^2 = F_\alpha F_\alpha^*$, which gives
\begin{equation}
    F_\alpha F_\alpha^* = \frac{1}{4(1+\hat\Omega\cdot\hat{p}_\alpha)^2} \left[ (\hat{m}\cdot\hat{p}_\alpha)^2 + (\hat{n}\cdot\hat{p}_\alpha)^2\right]^2.
\end{equation}
We then set up a coordinate system according to \autoref{fig:psr_gw_geometry}. We decompose the position vector of the pulsar into components that are parallel and perpendicular to $\hat\Omega$, such that $\bm{\hat{p}} = -\cos \xi_\alpha \, \bm{\hat{\Omega}} + \sin \xi_\alpha \, \bm{\hat{\rho}}$, where $\xi_\alpha$ is the angular separation between the pulsar and the GW source, and $\hat\rho$ is a vector lying in the same plane as $\hat{m}$ and $\hat{n}$. We let this \(\bm{\hat{\rho}} = \cos \varphi \, \bm\hat{m} + \sin \varphi \, \bm\hat{n}\), where $\varphi$ is an angle made by the pulsar’s position vector when it is projected into the plane perpendicular to $\bm{\hat{\Omega}}$. Doing so yields,
\begin{equation}
    F_\alpha F_\alpha^* = \frac{1}{4}(1+\cos\xi_\alpha)^2.
\end{equation}

Finally, we see
\begin{align}
    \overline{\mathrm{SNR}^2_\alpha} 
    =& \,\,\,\frac{4}{5}\frac{\mathcal{M}^{10/3}}{D_L^2 \omega^{2/3}}\frac{T}{\sigma_\alpha^2\Delta t} \nonumber\\ 
    & \times (1+\cos\xi_\alpha)^2 \sin^2\left[ \omega L_\alpha (1-\cos\xi_\alpha)\right].
\end{align}

\begin{figure} 
    \includegraphics[width=\columnwidth]{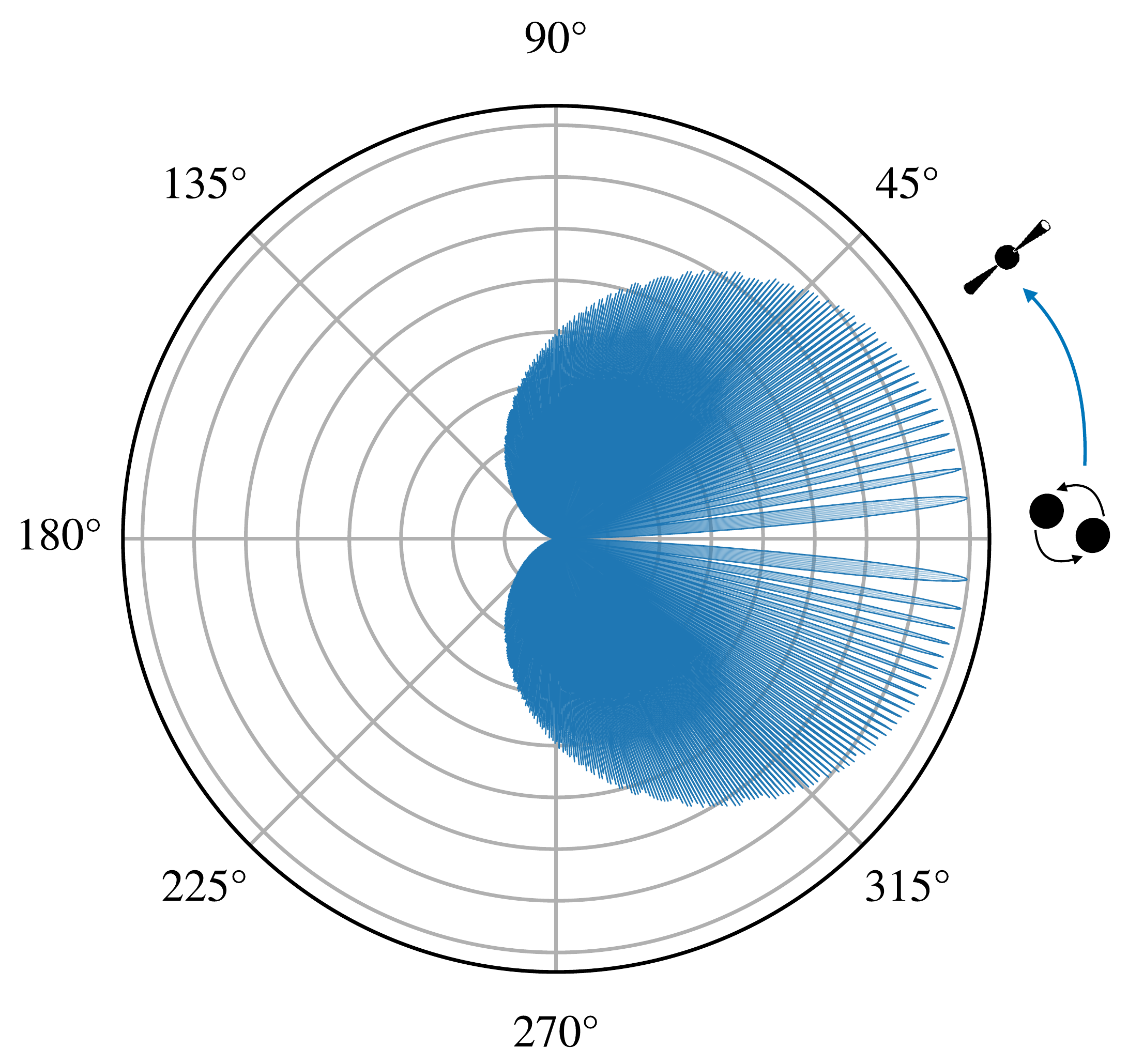}
    \caption{\label{fig:interference_viz}A polar-diagram visualization of the angular factors of $|\overline{\mathrm{SNR}}|\propto (1+\cos\xi)|\sin[\omega L (1-\cos\xi)]|$. The GW source position is kept fixed at $0^\circ$ while the pulsar is moved counter-clockwise to be at varying angles around the source. For visualization purposes, the pulsar distance has been reduced to $100$~pc. Inspired by a similar visualization in \citet{2011MNRAS.414.3251L}.}
\end{figure}

This again makes the interference between pulsar and Earth terms explicit, as shown in \autoref{fig:interference_viz}, following similar visualizations in \citet{2011MNRAS.414.3251L} and \citet{2024CQGra..41q5008R}. Pulsars that align precisely with the source position ($\xi_\alpha=0$) exhibit a net zero response to the GW signal due to destructive interference between the pulsar and Earth terms, such that $\overline{\mathrm{SNR}_\alpha}=0$; the radio pulses are said to be ``surfing'' on the GW, imparting no net signal. Similarly, pulsars that are on the opposite side of the sky to the GW source ($\xi_\alpha=\pi$) exhibit zero response due to the transverse nature of GWs, since $(\hat{m}\cdot\hat{p}_\alpha)=0$, $(\hat{n}\cdot\hat{p}_\alpha)=0$, and thus $F_\alpha=0$ there.

\section{GW signal localization} \label{sec:sec_localization}

\subsection{Fisher matrix uncertainty estimation}

The Fisher information matrix for a general data model can be written as the ensemble-averaged Hessian of the log-likelihood function evaluated around the true model parameters, $\theta_0$,
\begin{equation}
    \mathcal{I}_{ij} = -\langle \partial_i\partial_j \ln p(d|\theta_0)\rangle,
\end{equation}
where $d$ is data, and $\{i,j\}$ denote different signal parameters. Further specifying our model as a continuous GW signal in PTA data (where we only care about signal parameters) we write this in terms of the log-likelihood ratio,
\begin{equation}
    \ln \Lambda = (d|\mathbb{R}[s]) - \frac{1}{2}(\mathbb{R}[s]\,|\,\mathbb{R}[s]),
\end{equation}
such that
\begin{equation}
    \mathcal{I}_{ij} = (\partial_i \mathbb{R}[s]\, |\, \partial_j \mathbb{R}[s]).
\end{equation}

The Fisher matrix for the entire PTA can be written as a sum over per-pulsar Fisher matrices, $\mathcal{I}_{ij} = \sum_\alpha \mathcal{I}_{\alpha,ij}$, similar to adding per-pulsar SNRs in quadrature.

Via the Cauchy-Schwarz inequality, the inverse of the Fisher information matrix constitutes the Cram\'er-Rao (lower) variance bound on the unbiased estimation of parameters $\vec\theta$, such that
\begin{equation}
    C_{ij} \geq C^\mathrm{CR}_{ij} = [\mathcal{I}]^{-1}_{ij}.
\end{equation}

\subsection{Fisher sub-matrix for GW sky localization} \label{sec:fisher_submatrix_gwsky}

If we are only interested in a subset of model parameters — here, sky localization — we can marginalize over the uncertainties in the other nuisance parameters. We split the Fisher matrix into blocks of sky parameters, $\mathcal{I}^\mathrm{sky}$, nuisance parameters, $\mathcal{I}^{nn}$, and off-diagonal covariance blocks, $\mathcal{I}^{\mathrm{sn}}$, such that
\begin{equation}
    \mathcal{I} = \begin{pmatrix} 
                    \mathcal{I}^{\mathrm{sky}} &          \mathcal{I}^{\mathrm{sn},T} \\ \mathcal{I}^{\mathrm{sn}} & \mathcal{I}^{nn} 
                  \end{pmatrix}.
\end{equation}

Prior constraints on certain parameters (e.g., pulsar distances) can be enforced by adding a diagonal matrix of their inverse variances to $\mathcal{I}$. In the specific case of pulsar distances, which are considered nuisance parameters from the perspective of sky localization, the constraint matrix added to $\mathcal{I}$ would have $\left( 1/\sigma^2_{L,1}, 1/\sigma^2_{L,2}, \cdots, 1/\sigma^2_{L,N} \right)$ for the appropriate pulsar distance elements along the matrix diagonal, and zeros everywhere else. When we numerically examine the impact of pulsar distance precision on GW source localization later in \S\ref{sec:sec_numerical}, this is the procedure used.

The unmarginalized covariance on the sky parameters is given by $C^\mathrm{sky} = [\mathcal{I}^\mathrm{sky}]^{-1}$. This is instructive to consider and compute for analytic intuition, but it represents the unrealistic scenario in which we assume nuisance parameters are measured with infinite precision. By contrast, the marginalized covariance is given by taking the Schur complement of $\mathcal{I}^{nn}$, such that
\begin{align} \label{eq:marg_schur}
    \tilde{C}^\mathrm{sky} &\geq \tilde{C}^\mathrm{sky,CR} = (\tilde{\mathcal{I}}^{\mathrm{sky}})^{-1}, \\
    \tilde{\mathcal{I}}^\mathrm{sky} &= \mathcal{I}^\mathrm{sky} - [\mathcal{I}^\mathrm{sn}]^T [\mathcal{I}^{nn}]^{-1} \mathcal{I}^\mathrm{sn},
\end{align}
where $\mathcal{I}^{nn}$ is assumed to have relevant inverse variance constraints added. It is this marginalized measurement covariance of sky parameters that we will focus on in our numerical studies. We hereon suppress the Cram\'er-Rao (CR) superscript.

Following \citet{1998PhRvD..57.7089C} and \citet{2010PhRvD..81h2001W}, we compute the GW sky-localization area as
\begin{align} \label{eq:omega_marg}
    \Delta\Omega_\mathrm{sky} &= 2\pi\sqrt{\mathrm{det}[\tilde{C}^\mathrm{sky}]} \nonumber\\
                            &= \frac{2\pi}{\sqrt{\mathrm{det}[\tilde{\mathcal{I}}^{\mathrm{sky}}]}},
\end{align}
where, with this definition, the probability of a source lying \textit{outside} a solid angle $\Delta\Omega_0$ is $e^{-\Delta\Omega_0 / \Delta\Omega_\mathrm{sky}}$.

\subsection{Analytic Fisher localization scaling}

We now examine the angular partial derivatives needed to analytically compute the sky-coordinate Fisher sub-matrix, which is given by
\begin{align}
    \mathcal{I}_{ij}^{\text{sky}} &= \left( \partial_i \mathbb{R}[s]\, \middle|\, \partial_j \mathbb{R}[s] \right) \nonumber \\
    &\approx \frac{1}{2}\left( \partial_i s \middle| \partial_j s \right), \quad i,j \in \{ \cos\theta, \phi \},
\end{align}
where we again assume $T\gg1/\omega$ to keep our expressions in complex residual form.

Let us consider \autoref{eq:evol}. There are three parts of the residual model that have angular dependencies, and which influence PTA sky localization; $(i)$ the antenna response, $F$; $(ii)$ the pulsar term's phase, $\Phi_p$; and finally $(iii)$ the frequency of the pulsar term, $\omega_p$. \citet{2012PhRvD..86l4028B} considered the initial two in this list, but neglected the influence of the third. This is understandable since, as we shall see, the information derived from frequency evolution between the pulsar-term and Earth-term is sub-dominant to the other two effects, and only relevant for the most rapidly evolving binaries.

The result is a general equation for partial derivatives of the signal model with respect to sky coordinates,
\begin{equation}
    \partial_i s = \left(\frac{\partial_i F}{F}\right)s + \frac{\tilde{F}}{4i} \left[\, U\,\partial_i w -V\,\partial_i w^*\, \right],
\end{equation}
where
\begin{align}
    \partial_i w &= -\partial_i w_p \nonumber\\
                 &= w_p \left[ \frac{1}{3}\left(  \frac{\partial_i \omega_p}{\omega_p}\right) - 2i\,\partial_i\Phi_p\right]. 
\end{align}

For ease of analytic manipulation (without loss of generality), we can set $\psi=0$ and $\cos\iota=1$, such that $U=0$ and $V=4A$. Hence $s = i A F w^*$, and
\begin{equation} \label{eq:frac_s}
    \left( \frac{\partial_i s}{s} \right) = \left(\frac{\partial_i F}{F}\right) + \left(\frac{\partial_i w^*}{w^*}\right).
\end{equation}

We see that the three parts of the residual model encoding angular dependencies correspond to the following terms in $\partial_i s$: $(i)$ $(\partial_i F/F)$, $(ii)$ $(\partial_i \omega_p / \omega_p)$, and $(iii)$ $\partial_i\Phi_p$. The first term clearly encodes the fractional signal variation across the sky with respect to the antenna response geometry, while the second and third terms encode the influences of the pulsar term through its amplitude and phase modulations as a function of the directionally-dependent lag time, $\tau$.

We use the chain rule to change the partial derivatives of pulsar-term frequency and phase to partial derivatives of this lag time, $\tau$, such that
\begin{align}
    \partial_i \omega_p &= -\dot\omega_p\partial_i \tau, \\\nonumber
    \partial_i\Phi_p &= -[\omega_p + \dot\omega_p(t-t_0)]\partial_i \tau,
\end{align}
where the negative signs are because $\tau$ is a look-back time. Hence,
\begin{equation} \label{eq:partial_w}
    \partial_i w = -w_p\partial_i\tau\left\{ \frac{1}{3}\left(\frac{\dot\omega_p}{\omega_p}\right) - 2i[\omega_p + \dot\omega_p(t-t_0)]\right\}.
\end{equation}

\autoref{eq:frac_s} then becomes
\begin{align} \label{eq:partial_s_general}
    &\left( \frac{\partial_i s}{s} \right) = \\ \nonumber &\quad\left(\frac{\partial_i F}{F}\right) - \frac{w_p^* \,\,\partial_i\tau}{w_e^*-w_p^*} \left\{ \frac{1}{3}\left(\frac{\dot\omega_p}{\omega_p}\right) +  
    2i[\omega_p + \dot\omega_p(t-t_0)]\right\}.
\end{align}

Let us briefly examine this in the non-evolving signal scenario, where
\begin{align}
   w_p &= \frac{e^{2i\Phi} e^{-2i\omega\tau}}{\omega^{1/3}}, \nonumber\\
   \partial_i \Phi_p &= -\omega \partial_i \tau,
\end{align}

such that,

\begin{equation}
   \left( \frac{\partial_i w}{w} \right) = \frac{e^{-i\omega\tau}}{\sin(\omega\tau)}(\omega\partial_i \tau),
\end{equation}

and,

\begin{align} \label{eq:partial_s}
   \left( \frac{\partial_i s}{s} \right)
   &= \left(\frac{\partial_i F}{F}\right)
    + \left(\frac{\partial_i w^*}{w^*}\right) \nonumber\\
   &= \underbrace{\left(\frac{\partial_i F}{F}\right)}_{\text{antenna response}}
    + \underbrace{\frac{e^{i\omega\tau}}{\sin(\omega\tau)}(\omega\partial_i \tau)}_{\text{earth-pulsar--term interference}} .
\end{align}

We can draw several observations from \autoref{eq:partial_s} (and its more general antecedent, \autoref{eq:partial_s_general}). The first term contributes to GW source localization from angular derivatives of the antenna response functions, which vary slowly across the sky. 

By contrast, the second term gains much more precise localization information from the rapidly-oscillating phase interference of the Earth- and pulsar-terms, the pattern of which depends strongly on GW source location. We see that the denominator has zero points dictated by $\omega\tau=\omega L(1-\cos\xi)$, such that $\cos\xi=1-(n\pi/\omega L)$, where $n$ is an integer wrapping of the denominator term. When $\xi\ll 1$, the first oscillation of this interference term occurs at $\xi_1\sim\sqrt{2\pi/\omega L}$. With typical values of $\omega\sim \pi\times10^{-8}$~Hz and $L\sim1$~kpc, we have $\omega L\sim3\times10^3$, such that $\xi_1\sim0.04$~radians, or $\sim 2.5$~degrees from the GW source. A corollary to this inverse relationship between the interference fringe width and pulsar distance is that more distant pulsars would dominate a PTA's localization capabilities were their distances known sufficiently well \citep{2026A&A...706A.299G}.

The large magnitude of $\omega L$ also implies that the second term should be the dominant influence in the Fisher elements for sky location of the GW source (i.e., since $\omega\partial_i\tau = \omega L\,\partial_i(1+\hat\Omega\cdot\hat{p})$), with one important caveat. Where the first antenna response term only depends on the modeled GW source location and the well-known sky location of the pulsars, the second term relies on well-measured pulsar distances. In effect, we must know the pulsar distances to within one wrapping of this oscillatory term, such that $\sigma_L \leq\lambda_\mathrm{GW}/(1-\cos\xi)$. If they are not known to within this precision, then the major benefits of this interference term will be washed out when the pulsar distances constitute many of the nuisance parameters over which the sky block of the Fisher matrix is marginalized using the Schur complement; see \autoref{eq:marg_schur}. Practically, this will mostly leave the antenna response as the leverage for source localization.

We note that \autoref{eq:partial_s} describes the signal variation with respect to source position on the sky from antenna response geometry and the leading-order Earth-pulsar interference term under the non-evolving signal scenario; these correspond to the $\mathcal{B}$ and $\mathcal{A}$ signal derivative terms, respectively, from \citet{2012PhRvD..86l4028B}. However, this ignores several factors from \autoref{eq:partial_w} when binary evolution over $\tau$ is non-negligible. We will study the potential effect of these other factors in \S\ref{sec:precise_theory}.

We now examine the behavior of the sky-coordinate Fisher sub-matrix under the two scenarios where either pulsar distances are known imprecisely (i.e., the antenna response derivative term dominates in \autoref{eq:partial_s_general} since interference effects will wash out when pulsar distances are marginalized) or precisely (i.e., the interference term dominates in \autoref{eq:partial_s_general}).

\subsubsection{Imprecisely-measured pulsar distances} \label{sec:imprecise_theory}

\begin{figure} 
    \includegraphics[width=\columnwidth]{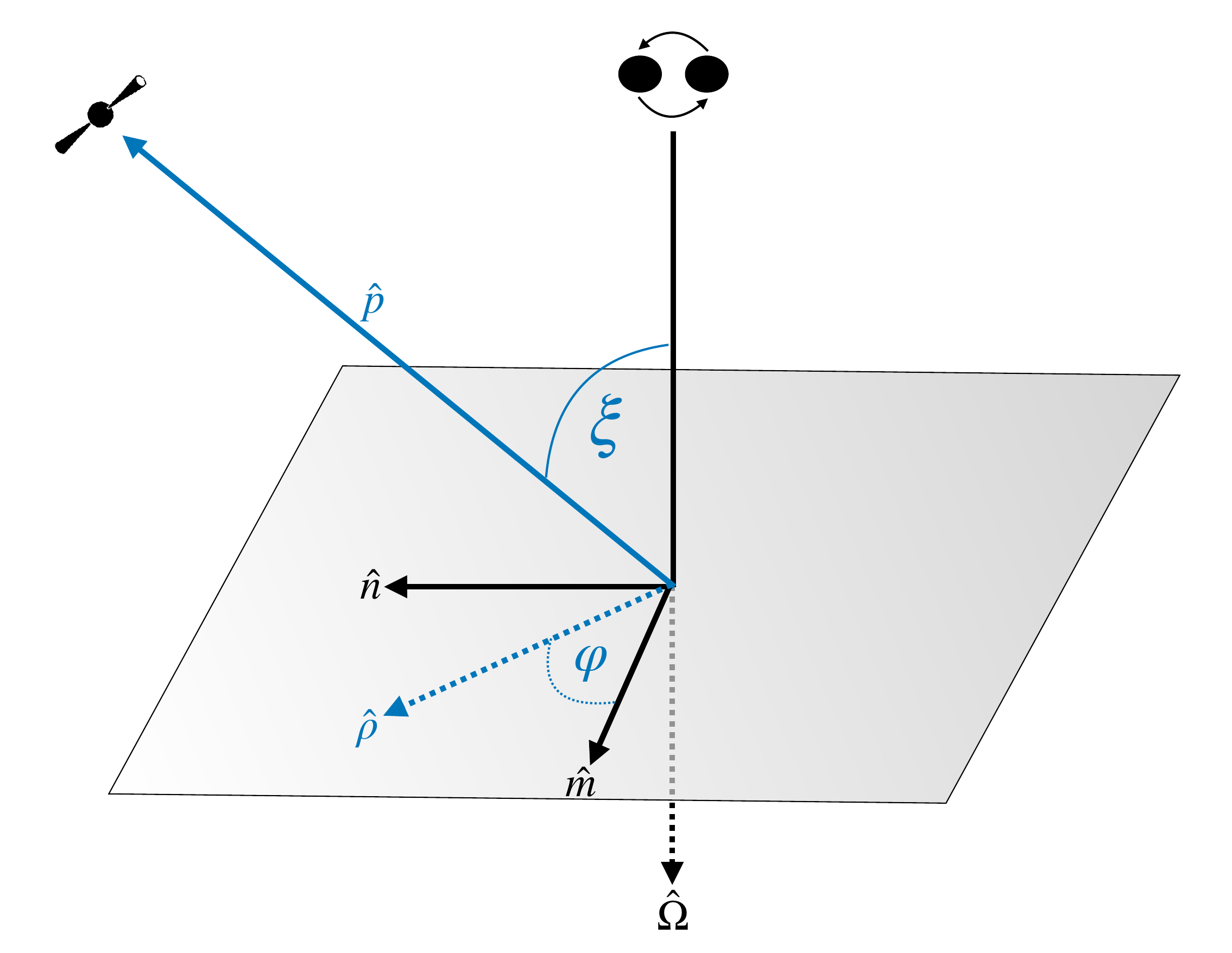}
    \caption{\label{fig:psr_gw_geometry}We employ a coordinate system in which the unit vector pointing from the Earth (or Solar System Barycenter) to a pulsar, $\hat{p}$, is decomposed into components that are parallel or perpendicular to the direction of GW propagation, $\hat\Omega$. The angle between the GW source and the pulsar is $\xi$. The projection of $\hat{p}$ into the plane transverse to $\hat\Omega$ is $\hat\rho$, which also contains $\hat{m}$ and $\hat{n}$, such that $\hat\Omega=\hat{m}\times\hat{n}$. The angle between $\hat\rho$ and $\hat{m}$ is $\varphi$.}
\end{figure}

In this scenario, the primary means by which GW signals are localized are the antenna response functions. Hence a Fisher uncertainty study requires taking the angular derivative of these response functions, the starting point of which is recalling \autoref{eq:complex_F}:
\begin{equation}
    F = \frac{1}{2} \frac{\left[ (\bm\hat{m} + i \bm\hat{n}) \cdot \bm{\hat{p}} \right]^2}{(1 + \bm{\hat{\Omega}} \cdot \bm{\hat{p}})} = F^{+} + i F^{\times}.
\end{equation}

Following \citet{2012PhRvD..86l4028B}, we use Rodrigues' rotation formula to define infinitesimal changes to $\bm{\hat{\Omega}}$ as rotations by an angle $\delta\gamma^{a}$ about the transverse basis vectors $\bm{\hat{m}}$ and $\bm{\hat{n}}$, i.e., $\delta\hat\Omega~=~(\hat{q}_{a}~\times~\hat\Omega)\delta\gamma^{_{a}}$, such that
\begin{equation}
    \partial_{q_{a}}\Omega = (\hat{q}_{a}\times\hat\Omega),
\end{equation}
where $\bm{q_{a}}~=~\{\bm\hat{m}, \bm\hat{n}\}$, with $a~=~(1,2)$.

Taking the partial derivative of $F$ with respect to sky angles then becomes
\begin{align} \label{eq:partial_F}
    \partial_{q_{a}} F = \frac{1}{ 2(1 + \bm{\hat{\Omega}} \cdot \bm{\hat{p}})^2 } &\left\{ (1 + \bm{\hat{\Omega}} \cdot \bm{\hat{p}}) \, \partial_{q_{a}} \left[ (\bm\hat{m} + i \bm\hat{n}) \cdot \bm{\hat{p}} \right]^2 \right.\nonumber\\
    &\left. - \left[ (\bm\hat{m} + i \bm\hat{n}) \cdot \bm{\hat{p}} \right]^2 \partial_{q_{a}} (1 + \bm{\hat{\Omega}} \cdot \bm{\hat{p}})  \right\},
\end{align}
where the numerator term, $\{\cdots\}$, in reduced form is,
\begin{align}
    \{\cdots\} =& 2 (1 + \bm{\hat{\Omega}} \cdot \bm{\hat{p}}) \left[ (\bm\hat{m} + i \bm\hat{n}) \cdot \bm{\hat{p}} \right] \left[ \partial_{q_{a}} (\bm\hat{m} \cdot \bm{\hat{p}}) + i \, \partial_{q_{a}} (\bm\hat{n} \cdot \bm{\hat{p}}) \right] \nonumber\\
    &- \left[ (\bm\hat{m} + i \bm\hat{n}) \cdot \bm{\hat{p}} \right]^2 \partial_{q_{a}} (\bm{\hat{\Omega}} \cdot \bm{\hat{p}}).
\end{align}
This numerator requires the following distinct partial derivatives:
\begin{align} \label{eq:omega_m_n_partials}
    \partial_{q_{a}} (\bm{\hat{m}} \cdot \bm{\hat{p}}) &= (\bm{q_{a}} \times \bm{\hat{m}}) \cdot \bm{\hat{p}} = (\bm\hat{m} \times \bm{\hat{p}}) \cdot \bm{q_{a}}, \nonumber\\
    \partial_{q_{a}} (\bm{\hat{n}} \cdot \bm{\hat{p}}) &= (\bm{q_{a}} \times \bm{\hat{n}}) \cdot \bm{\hat{p}} = (\bm\hat{n} \times \bm{\hat{p}}) \cdot \bm{q_{a}}, \nonumber\\
    \partial_{q_{a}} ( \bm{\hat{\Omega}} \cdot \bm{\hat{p}}) &= (\bm{q_{a}} \times \bm{\hat{\Omega}}) \cdot \bm{\hat{p}} = (\bm{\hat{\Omega}} \times \bm{\hat{p}}) \cdot \bm{q_{a}},
\end{align}
where in the final equality of each line we use the cyclic property of the scalar triple product. To make further progress, we re-use the angle projections introduced in \S\ref{sec:snr} and shown in \autoref{fig:psr_gw_geometry}, defining $\bm{\hat{p}} = -\cos \xi \, \bm{\hat{\Omega}} + \sin \xi \, \bm{\hat{\rho}}$, where $\xi$ is the angular separation between the pulsar and the GW source, and $\hat\rho$ is a vector lying in the same plane as $\hat{m}$ and $\hat{n}$. 
With the definition $\hat\Omega=\hat{m}\times\hat{n}$, and again using the cyclic property of the scalar triple product, we can now evaluate the partial derivatives from \autoref{eq:omega_m_n_partials} as
\begin{align}
    (\bm\hat{m} \times \bm{\hat{p}}) \cdot \bm{q_{a}} &= - \cos \xi (\bm\hat{m} \times \bm{\hat{\Omega}}) \cdot \bm{q_{a}} + \sin \xi (\bm\hat{m} \times \bm{\hat{\rho}}) \cdot \bm{q_{a}} \nonumber\\
    &= \cos \xi (\bm\hat{n} \cdot \bm{q_{a}}) = \delta_{a 2}\cos\xi \\
    (\bm\hat{n} \times \bm{\hat{p}}) \cdot \bm{q_{a}} &= - \cos \xi (\bm\hat{m} \cdot \bm{q_{a}}) = -\delta_{a 1}\cos\xi \\
    (\bm{\hat{\Omega}} \times \bm{\hat{p}}) \cdot \bm{q_{a}} &= \sin \xi (\bm{\hat{\Omega}} \times \bm{\hat{\rho}}) \cdot \bm{q_{a}}.
\end{align}

Finally, with \(\bm{\hat{\rho}} = \cos \varphi \, \bm\hat{m} + \sin \varphi \, \bm\hat{n}\) (where $\varphi$ is an angle made by the pulsar’s position vector when it is projected into the plane perpendicular to $\bm{\hat{\Omega}}$), all remaining scalar and vector products can be evaluated such that,
\begin{align}
    (1 + \bm{\hat{\Omega}} \cdot \bm{\hat{p}}) &= (1 - \cos \xi) \\
    (\bm\hat{m} + i \bm\hat{n}) \cdot \bm{\hat{p}} &= \sin \xi \left[ (\bm\hat{m} \cdot \bm{\hat{\rho}}) + i (\bm\hat{n} \cdot \bm{\hat{\rho}}) \right] \nonumber\\
    &= \sin \xi (\cos \varphi + i \sin \varphi) \nonumber\\
    &= \sin \xi \, e^{i\varphi},
\end{align}
which implies
\begin{equation}
    F = \frac{1}{2}(1+\cos\xi)e^{2i\varphi},
\end{equation}
and
\begin{align}
    (\bm{\hat{\Omega}} \times \bm{\hat{\rho}}) \cdot \bm{q_{a}} &= [\cos \varphi (\bm{\hat{\Omega}} \times \bm\hat{m}) + \sin \varphi (\bm{\hat{\Omega}} \times \bm\hat{n})] \cdot \bm{q_{a}} \nonumber\\
    &= [\cos \varphi \, \bm\hat{n} - \sin \varphi \, \bm\hat{m}] \cdot \bm{q_{a}} \nonumber\\
    &= \cos \varphi \, \delta_{a2} - \sin \varphi \, \delta_{a1}
\end{align}

With these simplifications, the numerator term, $\{\cdots\}$, of \autoref{eq:partial_F} becomes
\begin{align}
    \{\cdots\} =& \sin \xi \, e^{i\varphi} \left[ 2(1 - \cos \xi) \cos \xi (\delta_{a 2} - i \delta_{a 1}) \right. \nonumber\\
    &\left.- \sin^2 \xi \, e^{i\varphi} (\cos \varphi \, \delta_{a 2} - \sin \varphi \, \delta_{a 1}) \right] \nonumber\\
    =& \left[ \sin^3 \xi \, e^{2i\varphi} \sin \varphi - i \sin 2\xi (1 - \cos \xi) e^{i\varphi} \right] \delta_{a 1} \nonumber\\
    &+ \left[ \sin 2\xi (1 - \cos \xi) e^{i\varphi} - \sin^3 \xi \, e^{2i\varphi} \cos \varphi \right] \delta_{a 2}.
\end{align}

Using this result and its relation to $\partial_i s$, we can now write the time-integrated, polarization-averaged, and inclination-averaged elements of the sky-coordinate Fisher sub-matrix as
\begin{align}
    \overline{\mathcal{I}^\mathrm{sky}_{ij}} &\approx \frac{1}{2} \left( \partial_i s | \partial_j s \right) \nonumber\\
    &= \overline{\mathrm{SNR}^2} \times\frac{(\partial_i F)(\partial_j F^*)}{FF^*} \nonumber\\
    &= \overline{\mathrm{SNR}^2} \times\frac{(\partial_i F^+)(\partial_j F^+) + (\partial_i F^\times)(\partial_j F^\times)}{(F^+)^2 + (F^\times)^2}.
\end{align}

Now averaging over the azimuthal projection angle $\varphi\in U[0,2\pi]$, and taking the square root of the determinant of this sky-coordinate sub-matrix, gives,
\begin{align}
    \mathrm{det}(\overline{\mathcal{I}^\mathrm{sky}})^{1/2} &= \overline{\mathrm{SNR}^2} \times \frac{\frac{1}{8} \cot^2 \left( \xi/2 \right) \left( 5 \cos^2 \xi - 2 \cos \xi + 1 \right)}{\frac{1}{4} (1+\cos\xi)^2} \nonumber\\
    &= \overline{\mathrm{SNR}^2} \times\frac{\left( 5 \cos^2 \xi - 2 \cos \xi + 1 \right)}{2\sin^2\xi}.
\end{align}

Therefore, an analytic approximation of the scaling behavior for antenna-response--driven GW localization precision is,
\begin{equation} \label{eq:analytic_response_omega}
    \Delta\Omega_\mathrm{sky} \gtrapprox \left\{ \displaystyle\sum_\alpha \frac{\overline{\mathrm{SNR_\alpha}^2}}{4\pi}  \times\frac{\left( 5 \cos^2 \xi_\alpha - 2 \cos \xi_\alpha + 1 \right)}{\sin^2\xi_\alpha}\right\}^{-1}, 
\end{equation}
which, when $\xi\ll 1$, becomes
\begin{equation} \label{eq:imprecise_scaling}
    \Delta\Omega_\mathrm{sky} \gtrapprox \left\{ \displaystyle\sum_\alpha \frac{\overline{\mathrm{SNR_\alpha}^2}}{\pi\,\xi_\alpha^2} \right\}^{-1}, 
\end{equation}
in agreement with the small-angle scaling behavior of \citet{2012PhRvD..86l4028B}.

\subsection{Well-measured pulsar distances} \label{sec:precise_theory}

The situation is actually rather simpler when interference between the Earth- and pulsar-term dominates the signal derivative with respect to sky coordinates. We begin with,
\begin{align} 
    \left( \frac{\partial_i s}{s} \right) \approx - \frac{w_p^* \,\,\partial_i\tau}{w_e^*-w_p^*} \left\{ \frac{1}{3}\left(\frac{\dot\omega_p}{\omega_p}\right) +  
    2i[\omega_p + \dot\omega_p(t-t_0)]\right\},
\end{align}
which shows that we must evaluate the partial derivative of the lag time, which is,
\begin{align} \label{eq:partial_interf}
    \partial_{q_{a}} \tau &= L\,\partial_{q_{a}}(1+\hat\Omega\cdot\hat{p}) \nonumber\\
    &= L\,\sin\xi\, (\hat\Omega\times\hat\rho)\cdot q_{a} \nonumber\\
    &= L\,\sin\xi\,\left[\cos \varphi \, \delta_{a 2} - \sin \varphi \, \delta_{a 1} \right],
\end{align}
where in the second and third lines we have re-used relationships from \S\ref{sec:imprecise_theory}. 

In this scenario, the time-integrated, polarization-averaged, and inclination-averaged elements of the sky-coordinate Fisher sub-matrix are given by
\begin{align}
    \overline{\mathcal{I}^\mathrm{sky}_{ij}} &\approx \frac{1}{2} \left( \partial_i s | \partial_j s \right) \nonumber\\
    &\approx \overline{\mathrm{SNR}^2} \times (\omega_p\partial_i\tau)(\omega_p\partial_j\tau) \\ \nonumber
    &\quad\times \frac{4|w_p|^2}{\overline{|w_e-w_p|^2}} \overline{\left\{ \left[1 + \frac{\dot\omega_p}{\omega_p}(t-t_0)\right]^2 + \frac{1}{36}\left(\frac{\dot\omega_p}{\omega_p^2}\right)^2\right\}},
\end{align}
where $|w_p|^2 = \omega_p^{-2/3}$, and the time-integrated quantities (or time-averaged, with the same result, since they appear in a ratio) are,
\begin{align}
    \overline{|w_e-w_p|^2} &= \frac{1}{T}\int_{t_0-T/2}^{t_0+T/2} dt \left[\omega^{-2/3} + \omega_p^{-2/3} \right. \nonumber \\ 
    &\quad\qquad \left.- 2(\omega\omega_p)^{-1/3}\cos(2\Delta\Phi) \right] \nonumber \\
    &= \quad\omega^{-2/3} + \omega_p^{-2/3} \nonumber\\
    &\quad- 2(\omega\omega_p)^{-1/3}\cos(2\Delta\Phi_0)\sinc(\Delta\omega T)
\end{align}
where $\Delta\Phi = \Phi - \Phi_p = \Phi_0-\Phi_{p,0} + \omega(t-t_0) - \omega_p(t-t_0) = \Delta\Phi_0 + \Delta\omega (t-t_0)$, and
\begin{align}
    &\overline{\left\{ \left[1 + \frac{\dot\omega_p}{\omega_p}(t-t_0)\right]^2 + \frac{1}{36}\left(\frac{\dot\omega_p}{\omega_p^2}\right)^2\right\}} \nonumber\\
    &\qquad\qquad\qquad= 1 + \frac{1}{12}\left(\frac{\dot\omega_p T}{\omega_p}\right)^2 + \frac{1}{36}\left(\frac{\dot\omega_p}{\omega_p^2}\right)^2.
\end{align}

Hence,
\begin{align}
    \overline{\mathcal{I}^\mathrm{sky}_{ij}} &\approx \frac{1}{2} \left( \partial_i s | \partial_j s \right) \nonumber\\
    &\approx 4\,\overline{\mathrm{SNR}^2} \times (\omega_p\partial_i\tau)(\omega_p\partial_j\tau) \\ \nonumber
    &\times \frac{1 + \frac{1}{12}(\dot\omega_p T/\omega_p)^2 + \frac{1}{36}(\dot\omega_p/\omega_p^2)^2}{1 + (\omega_p/\omega)^{2/3} - 2(\omega_p/\omega)^{1/3}\cos(2\Delta\Phi_0)\sinc(\Delta\omega T)} \nonumber\\
    &\approx 4\,\overline{\mathrm{SNR}^2} \times (\omega_p\partial_i\tau)(\omega_p\partial_j\tau) \times\frac{\mathcal{N}}{\mathcal{D}}.
\end{align}

Now substituting \autoref{eq:partial_interf}, averaging over the azimuthal projection angle $\varphi\in U[0,2\pi]$, and taking the square root of the determinant of this sky-coordinate sub-matrix, gives,
\begin{equation} \label{eq:root_det_interference}
    \mathrm{det}(\overline{\mathcal{I}^\mathrm{sky}})^{1/2} = 2\,\overline{\mathrm{SNR}^2} \times (\omega_p L)^2 \sin^2\xi \times \frac{\mathcal{N}}{\mathcal{D}}.
\end{equation}

Therefore, an analytic approximation of the scaling behavior for interference-driven GW localization precision is,
\begin{align} \label{eq:analytic_interference_omega}
    &\Delta\Omega_\mathrm{sky} \gtrapprox \left\{ \displaystyle\sum_\alpha \overline{\mathrm{SNR_\alpha}^2}  \times \frac{(\omega_{p,\alpha} L_\alpha)^2 \sin^2\xi_\alpha}{\pi} \times \frac{\mathcal{N}_\alpha}{\mathcal{D}_\alpha}\right\}^{-1}.
\end{align}

The key term in \autoref{eq:analytic_interference_omega} that drives GW localization precision is $\mathcal{D}$, and in particular $\cos(2\Delta\Phi_0)\sinc(\Delta\omega T)$. This encodes the influence of phase differences between the Earth-term and pulsar-term that oscillate rapidly on the sky, as well as frequency evolution between the terms that manifests as a beat envelope.

\autoref{eq:analytic_interference_omega} significantly simplifies in the absence of evolution, such that $\omega_p \approx \omega$ and $\Phi_p = \Phi - \omega\tau$, in which case $\mathcal{N} = 1$ and $\mathcal{D} = 2 - 2\cos(2\omega\tau) = 4\sin^2(\omega\tau)$. Hence,
\begin{equation} \label{eq:analytic_interference_omega_nevolve}
   \Delta\Omega_\mathrm{sky} \gtrapprox \left\{ \displaystyle\sum_\alpha \frac{\overline{\mathrm{SNR_\alpha}^2}}{4\pi}  \frac{(\omega L_\alpha)^2 \sin^2\xi_\alpha}{\sin^2 [\omega L_\alpha (1-\cos\xi_\alpha)]} \right\}^{-1}.
\end{equation}
which, when $\xi\ll 1$, becomes
\begin{equation} \label{eq:precise_scaling}
    \Delta\Omega_\mathrm{sky} \gtrapprox \left\{ \displaystyle\sum_\alpha \frac{\overline{\mathrm{SNR_\alpha}^2}}{\pi \xi_\alpha^2}  \frac{1}{\mathrm{sinc}^2(\omega L_\alpha\xi^2_\alpha/2)} \right\}^{-1}.
\end{equation}

We see that \autoref{eq:precise_scaling} has similar leading-order scaling as \autoref{eq:imprecise_scaling} in the limit of small angular separation between the pulsar and GW source, modified by the $\mathrm{sinc^2}(\cdot)$ interference term and the binary evolution term. 

As a parting note to this section, both \autoref{eq:analytic_interference_omega} and \autoref{eq:analytic_response_omega} are derived under the Fisher approximation, and do not marginalize over uncertainties in other signal parameters. Real GW localization precision will be worse than these scaling relationships imply.

\section{Numerical Studies} \label{sec:sec_numerical}

\begin{figure} 
    \includegraphics[width=\columnwidth]{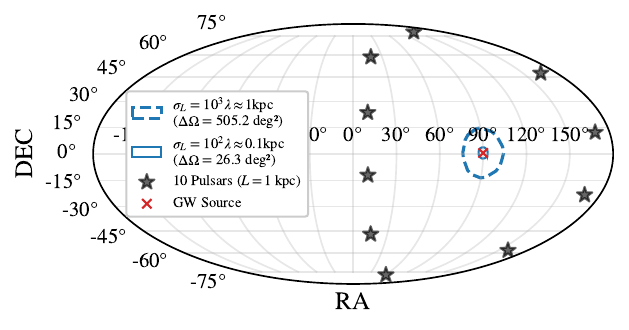}
    \caption{\label{fig:study0}A circular ring of $10\times1$-$\mathrm{kpc}$ pulsars (black stars) are arranged around a GW source (red cross) at $(\cos\theta=0,\phi=\pi/2)$. We visualize the marginalized GW localization precision in terms of ellipses around the source, doing so for two different precisions on pulsar distances: $\sigma_L=10^3\lambda_\mathrm{GW}\approx1\,\mathrm{kpc}$ (blue dashed), and $\sigma_L=10^2\lambda_\mathrm{GW}\approx0.1\,\mathrm{kpc}$ (blue solid).}
\end{figure}

The previous sections have laid out the theoretical formalism for GW source localization with PTAs, and the analytical Fisher results have provided some intuition for the key drivers. In this section, we confront this intuition with numerical studies that do not rely on many of the approximations needed for earlier analytic headway. Our numerical studies will still be somewhat idealized: we ignore pulsar timing models, (a)chromatic noise effects, and we will mostly examine arrays of equidistant pulsars in circular rings around a GW source. 

The properties of our GW source are chosen to be typical of what we may expect of a future detection, i.e., $f_\mathrm{GW}=\omega/\pi=10~\mathrm{nHz}$ and $\mathcal{M}=10^{9.5}M_\odot$. We keep the binary face-on ($\cos\iota=1$), with $\psi=0$, and $\Phi_0=0$. For the PTA, we assume identical pulsars in all but their sky location, observed $100$ times over a baseline of $10$~years with $100~\mathrm{ns}$ TOA uncertainties. Unless otherwise specified, the GW source is located at $(\cos\theta=0,\phi=0)$ and pulsar distances are set to be $L=1~\mathrm{kpc}$. The GW source distance is calibrated to maintain $\mathrm{SNR}=10$.

The GW signal model is general, and does not assume weak or non-evolution of the binary's orbital frequency back to the pulsar term. However we do assume that binary evolution over the observational baseline is negligible. All signal components are written in \texttt{JAX} [originally based on code from \citet{2024PhRvD.110f3038F}\footnote{\href{https://github.com/gabefreedman/etudes}{https://github.com/gabefreedman/etudes}}] to make use of automatic differentiation for partial derivatives, and speed gains through Just-In-Time (\texttt{JIT}) compilation. In all cases, we compute the Fisher matrix that is marginalized over nuisance parameters (i.e., all parameters except GW sky coordinates) and find $\Delta\Omega_\mathrm{sky}$ according to \autoref{eq:omega_marg}.

Example GW localization ellipses are shown in \autoref{fig:study0}, where the GW source is at $(\cos\theta=0,\phi=\pi/2)$ and $10$ pulsars are distributed in a ring of radius $80^\circ$ around the source. All pulsars are at $L=1~\mathrm{kpc}$ with uncertainties that are either $10^3$ or $10^2$ GW wavelengths, i.e., $\sigma_L\sim1$~kpc or $\sigma_L\sim0.1$~kpc, respectively. This change in pulsar distance uncertainty causes a factor of $\gtrsim20$ change in localization precision, despite the SNR being the same for both scenarios. We now study such effects systematically.

\subsection{Localization precision as a function of pulsar proximity to the GW source}

\begin{figure} 
\centering    \includegraphics[width=\columnwidth]{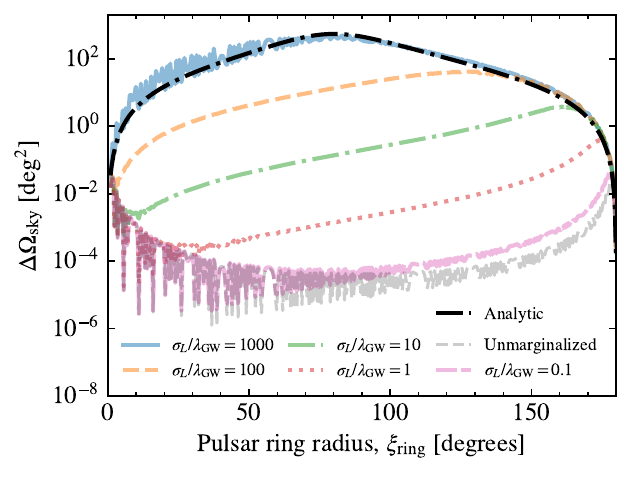}
    \caption{\label{fig:study1a}We systematically move our ring of $20\times 1$-$\mathrm{kpc}$ pulsars through varying angular radii to explore the influence on marginalized GW source localization precision. We show the impact of different precisions on pulsar distances as a fraction of the GW wavelength, and contrast with the unmarginalized localization precision (grey) and the form for the analytic antenna-response--driven localization (black dash-dot) given by \autoref{eq:analytic_response_omega}.}
\end{figure}

\autoref{fig:study1a} shows the marginalized GW source localization precision, $\Delta\Omega_\mathrm{sky}$, as a function of the radius of a $20$ pulsar ring array; this radius acts as a proxy for considering the importance of pulsar proximity to GW source localization. We consider five different values for the precision with which pulsar distances are known: $\sigma_L/\lambda_\mathrm{GW}=[0.1,1,10,100,1000]$. For context, we also show the localization precision computed from solely the sky-coordinate Fisher sub-matrix without marginalization over nuisance variables, such as the other binary parameters or pulsar distances.

We notice a clear transition of the key driving influence of GW source localization from Earth-pulsar--term interference when pulsar distances are known precisely ($\sigma_L/\lambda_\mathrm{GW}=0.1$), to antenna response modulations when pulsar distances are known imprecisely ($\sigma_L/\lambda_\mathrm{GW}=1000$). In the former scenario, the overall behavior reaches best localization precision at $\sim90^\circ$, in agreement with our intuition from \autoref{eq:root_det_interference}. The rapid fluctuations we observe in \autoref{fig:study1a} are not numerical instabilities, but rather the direct result of interference via the rapidly-varying $\sin(\omega\tau)=\sin[\omega L (1-\cos\xi)]$ term. We also see that the analytic scaling relationship for antenna-response--driven localization given by \autoref{eq:analytic_response_omega} agrees remarkably well with numerical results in the case of imprecisely-measured pulsar distances ($\sigma_L/\lambda_\mathrm{GW}=1000$). 

The case $\sigma_L/\lambda_\mathrm{GW}=0.1$ gets very close to the unmarginalized idealized scenario. However, even by the time we get to $\sigma_L/\lambda_\mathrm{GW}=1$, the leverage we get from interference effects is starting to be washed out when we marginalize over pulsar distances, and the effect of antenna response modulations begins to influence. Knowing pulsar distances to the precision of a GW wavelength corresponds to a parsec-level measurement here, which is extraordinary, and only possible with J$0437$$-$$4715$ thus far \citep{2024ApJ...971L..18R}, although J$1909$$-$$3744$ and J$0030$$+$$0451$ are getting usefully close \citep{2021MNRAS.507.2137R,2023MNRAS.519.4982D}. We see that, for all intents and purposes, even when interference effects remain highly influential to localization, the proximity of pulsars to a GW source on the sky is critical. 

\begin{figure*} 
    \includegraphics[width=\columnwidth]{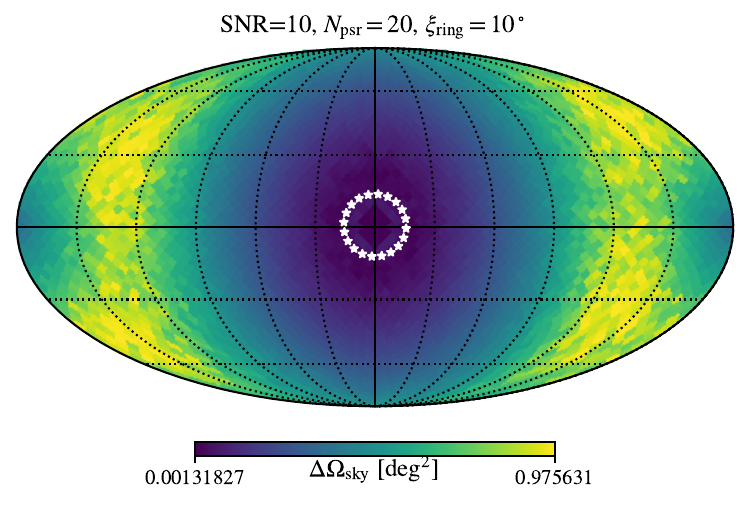}
    \includegraphics[width=\columnwidth]{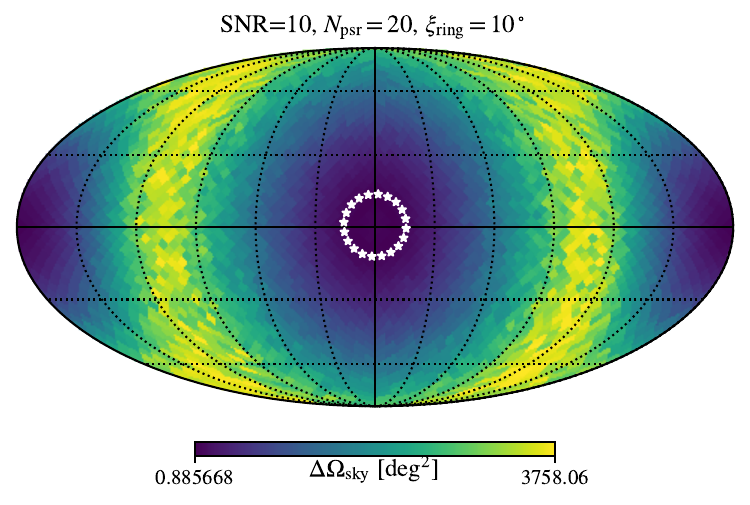}
    \caption{\label{fig:study2}We keep the angular radius of the pulsar ring fixed at $\xi_\mathrm{ring}=10^\circ$ and move the GW source around the sky. The colorbar shows the marginalized GW localization precision in square degrees, with the precision at the $1\,\mathrm{kpc}$ pulsar distances given by $1$~pc \textit{(left)} and $1$~kpc \textit{(right)}.}
\end{figure*}

This is reinforced when we vary our study to keep the pulsar ring radius fixed at $10^\circ$, and move the GW source around the sky instead of remaining at the ring center. \autoref{fig:study2} shows two sky maps that are colored by the precision with which our ring of pulsars can measure the location of a GW source in each pixel. The scale of the colorbar in each panel is very different, which is a direct consequence of the differing assumed pulsar distance precisions: $1$~pc \textit{(left)} and $1$~kpc \textit{(right)}. Even in the former case of precisely measured pulsar distances, the best GW source localization occurs within and in close proximity to the ring of pulsars.  

\subsection{Influence of pulsar distance and its uncertainty}

\begin{figure*} 
    \includegraphics[width=\columnwidth]{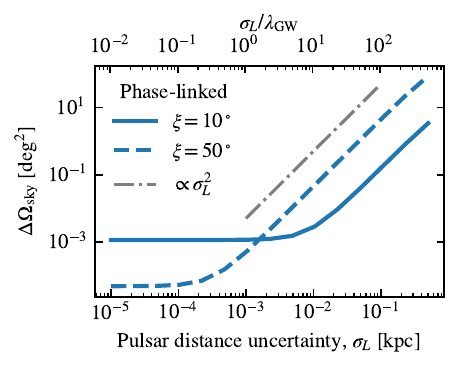}
    \includegraphics[width=\columnwidth]{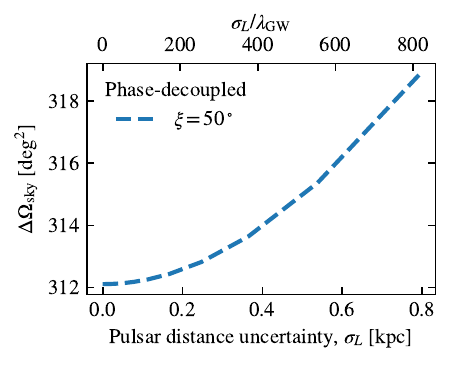}
    \caption{\label{fig:study4-5}We keep the GW source position and $20$-pulsar-ring radius fixed, then vary the precision with which the $1\,\mathrm{kpc}$-distant pulsars are known. We study this under the assumption that the pulsar-term phases are linked to the Earth-term phase through geometry-dependent shifts \textit{(left)}, or are simply treated as decoupled nuisance variables that are marginalized over \textit{(right)}. Note that the left panel has both axes in logarithmic space, while the right panel is linear; this is the cleanest visualization of the disparity between these approaches.}
\end{figure*}

Given the apparent impact of the precision with which we know pulsar distances on GW source localization, we study this effect more systematically at two fixed pulsar ring radii. In \autoref{fig:study4-5}, we keep the position of the GW source fixed at the center of $20\times1$-kpc pulsar rings that are either $\xi_\mathrm{ring}=10^\circ$ or $50^\circ$.

The left panel shows the kind of search scenario we have discussed thus far: the pulsar-term frequency and initial phase are entirely derived from the GW source properties (e.g., binary chirp mass, Earth-term orbital frequency, and phase), the pulsar distance, and the angular separation between the source and the pulsar. We call this the \textit{phase-linked} case. We see that the localization precision scales approximately with the square of the pulsar distance precision over a large range of values. This scaling reflects our enforcement of prior pulsar distance constraints in the Fisher analysis.

However, this scaling eventually saturates as we improve the distance precision, until no further localization improvement is possible. This saturation is reached when the PTA has reached its diffraction-limited localization \citep{2012PhRvD..86l4028B,2025arXiv251210795T}. This occurs at different distance precisions for the $10^\circ$ and $50^\circ$ pulsar ring radius cases. Since the key factor in assessing whether a pulsar distance is known precisely for PTA GW localization is $\sigma_L\leq\lambda_\mathrm{GW}/(1-\cos\xi)$, this diffraction limit is reached sooner (i.e., at worse pulsar distance precisions) when pulsars are in closer proximity to the GW source. The diffraction-limited localization precision is also worse for $\xi_\mathrm{ring}=10^\circ$ than $50^\circ$ because the Fisher uncertainty is dominated by interference effects, such that better localization precision is given when pulsars are further from the GW source (up to $\sim90^\circ$, then precision worsens again).

The right panel of \autoref{fig:study4-5} requires some important discussion. In this \textit{phase-decoupled} scenario, we do not attempt to describe the initial pulsar-term phase in terms of GW source properties, the pulsar distance, and the angular separation between the source and the pulsar. Rather, it is treated as a nuisance variable that we would intend to fit during parameter estimation and marginalize over. The only impact the pulsar distance and its measurement precision have on GW source localization is through the frequency of the pulsar term. In the language of \S\ref{sec:fisher_submatrix_gwsky}, we neglect information from $\partial_i\Phi_{p,0}$ such that $\partial_i\Phi_p = -\dot\omega_p(t-t_0)\partial_i\tau$. That leaves the numerator in this phase-decoupled scenario to be 
\begin{equation}
    \mathcal{N}_\mathrm{PD} = (\dot\omega_p/\omega_p^2)^2 [3(\omega_p T)^2 + 1]/36,
\end{equation}
which is $\sim10^{-8}$ for the fiducial binary values in this paper, and thus $10^8$ times worse in localization precision than the phase-linked scenario. This severely limits the impact that the pulsar term can have on GW localization precision. As we see in \autoref{fig:study4-5}, improving the pulsar distance precision by more than four orders of magnitude only improves GW sky localization precision by $\sim2\%$.

Why study such a scenario when it clearly throws so much valuable information away? Perhaps so, but as mentioned in the introduction, this is in fact how every single production-level PTA continuous-wave search pipeline currently functions \citep{2022PhRvD.105l2003B,2025PhRvD.112h3035G}. Early studies and prototypes of continuous GW search pipelines for PTAs found that rapid interference oscillations led to a vast number of secondary maxima in the likelihood landscape for GW source location and pulsar distances, rendering MCMC sampling extremely challenging \citep{2010arXiv1008.1782C}. While some novel solutions exist \citep{neil_2011_presentation,2013CQGra..30v4004E}, the community deemed it not worth the bother given the poor pulsar distance precisions at the time. Hence, the right panel of \autoref{fig:study4-5} represents the impact of pulsar distance precisions on GW source localization using status-quo search pipelines. 

\begin{figure} 
    \includegraphics[width=\columnwidth]{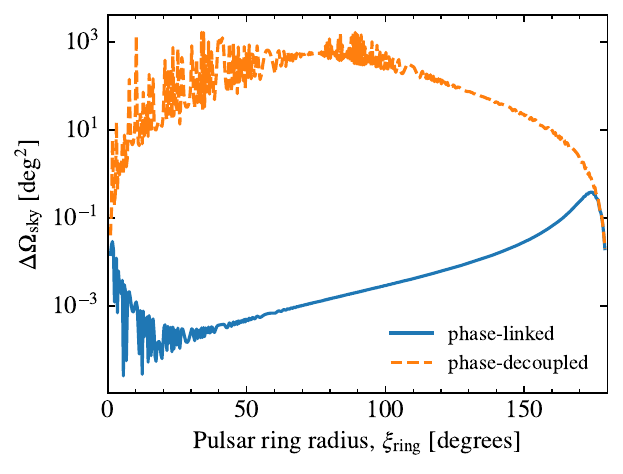}
    \caption{\label{fig:study3}We reiterate the study in \autoref{fig:study1a} for $\sigma_L/\lambda_\mathrm{GW}\approx1$ to compare the performances of the phase-linked and phase-decoupled approaches. By treating the pulsar-term phase as a nuisance variable over which we marginalize, phase-decoupled GW source localization is driven by antenna response information.}
\end{figure}

The performance of the phase-decoupled approach is compared with the phase-linked approach as a function of the radius of a $20\times1$-kpc ring array in \autoref{fig:study3}. We set $\sigma_L=1$~pc such that $\sigma_L/\lambda_\mathrm{GW}\approx1$. We see that the behavior of the phase-decoupled approach is akin to the most imprecisely-measured pulsar distance case ($\sigma_L/\lambda_\mathrm{GW}=10^3$) of the phase-linked approach in \autoref{fig:study1a}. Hence, the general behavior of the phase-decoupled approach is that of antenna-response--driven GW localization, which is best when pulsars are in close proximity to the source.

\begin{figure} 
    \includegraphics[width=\columnwidth]{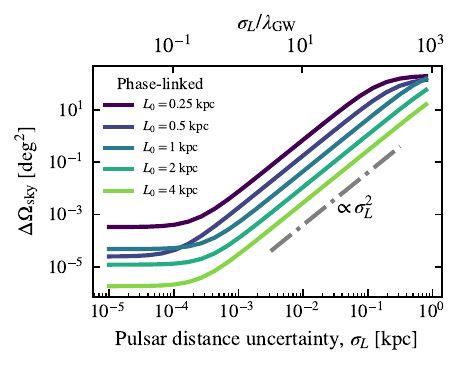}
    \caption{\label{fig:varied_pdist}We vary the absolute distances to $20$ pulsars in a $50^\circ$-radius ring around a GW source. The GW localization precision scales in a similar way to \autoref{fig:study4-5} as the pulsar distance precision is varied, with a prior-dominated portion at imprecise distance values that levels to a diffraction-limited floor. When correcting for interference modulations between the Earth and pulsar terms, GW localization improves with increasing pulsar distance as $1/L^2$.}
\end{figure}

Finally, we examine the influence of the absolute value of a pulsar's distance on GW localization precision. We assume, as before, that all $20$ pulsars have the same distance, which is varied between $0.25-4$~kpc. The result for the phase-linked approach, and a ring radius of $\xi_\mathrm{ring}=50^\circ$, is shown in \autoref{fig:varied_pdist}, where the precision with which these distances are known is also varied, as in \autoref{fig:study4-5}. Similarly to \citet{2026A&A...706A.299G}, more distant pulsars are observed to improve localization precision. While the variation with pulsar distance is seemingly non-monotonic, this is caused by interference --- the factor $\sin^2 [\omega L (1-\cos\xi)]$ from \autoref{eq:analytic_interference_omega} --- between the Earth term and pulsar terms. When corrected for this, the diffraction-limited floor on localization precision scales as $\Delta\Omega_\mathrm{sky}\propto 1/(\omega L)^2$, while the prior-dominated behavior scales, as before, like $\Delta\Omega_\mathrm{sky}\propto (\sigma_L/L)^2$. 

The transition between the prior-dominated region and the diffraction-limited floor remains independent of the absolute pulsar distance value, and determined rather by the criterion $\sigma_L\leq\lambda_\mathrm{GW}/(1-\cos\xi)$. By contrast, pulsar distance has negligible impact on localization precision in the phase-decoupled scenario, other than to produce non-monotonic variations $\sim120-320$~deg$^\circ$ caused by unmodeled interference effects between the Earth and pulsar terms.

\subsection{Influence of binary chirp mass}

\begin{figure} 
    \includegraphics[width=\columnwidth]{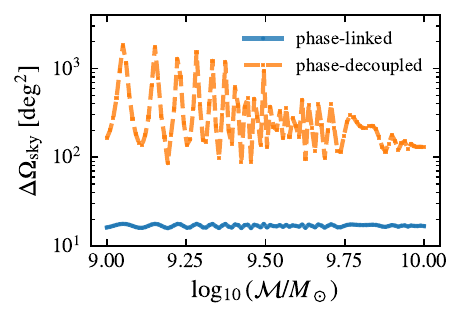}
    \caption{\label{fig:study5b}We investigate GW source localization precision as a function of source chirp mass under the phase-linked (blue) and phase-decoupled (orange) scenarios. The radius of our circular ring of $20\times1$-kpc pulsars is kept fixed at $\xi_\mathrm{ring}=50^\circ$, with $\sigma_L/\lambda_\mathrm{GW}=205$, i.e., $\sigma_L=0.2$ kpc.}
\end{figure}

As a final assessment of the difference between the phase-linked and phase-decoupled performances for localization precision, we systematically vary the chirp mass of the binary system, while scaling the binary distance to retain $\mathrm{SNR}=10$. We keep our $20\times1$-kpc pulsar ring radius fixed at $\xi_\mathrm{ring}=50^\circ$, with $\sigma_L/\lambda_\mathrm{GW}\approx 205$, i.e., $\sigma_L=0.2$~kpc.

\autoref{fig:study5b} shows the results, where the phase-linked analysis exhibits small oscillations due to the changing signal structure. The increasing chirp mass lowers the pulsar-term frequencies and affects the pulsar-term phases. Since the SNR is kept fixed and all pulsar-term information is being leveraged, increasing the chirp mass by an order of magnitude has a negligible impact on GW localization precision under the phase-linked scenario.

A much more significant impact is seen for the phase-decoupled scenario. No connection between the pulsar-term phases and binary/PTA properties is being accounted for, and hence the increasing chirp mass (with its consequent rapid oscillatory changes to pulsar-term phases) leads to pronounced constructive and destructive interference effects between the Earth and pulsar terms. A similar effect is observed in Figure 7 of \citet{2024ApJ...976..129P}, where changes in binary chirp mass lead to non-monotonic variations in GW localization precision (albeit there using a production-level Bayesian analysis pipeline). However, we notice that the severity of these interference oscillations diminishes at higher chirp masses. As the chirp mass increases, the pulsar-term frequencies lower, reducing the constructive and destructive interference effects and instead producing beat patterns between the two distinct signal frequencies in a given pulsar's time series. This separation between the Earth and pulsar terms damps the localization variations under the phase-decoupled scenario. A similar effect is seen in \citet{2026A&A...706A.299G}, where greater separation between the Earth and pulsar terms ---achieved in their case through more distant pulsars --- can reduce sky location biases and improve overall parameter recovery.

\section{Conclusions} \label{sec:sec_conclusions}

We have investigated the physics influencing GW source localization using pulsar timing arrays under different assumptions for the precision with which pulsar distances are known, their absolute distance values, and the angular proximity of pulsars to the source. We have provided a clear delineation of the signal factors that influence localization under the Fisher uncertainty approximation, identifying three key aspects: antenna response variations, pulsar-term phase modulations, and frequency evolution between the Earth and pulsar terms, where the latter provides the weakest information.

Our framework provides analytic intuition for the behavior of GW localization when pulsar distances are known precisely--- $\sigma_L\leq\lambda_\mathrm{GW}/(1-\cos\xi)$--- or imprecisely. In the former, interference between the Earth and pulsar term phases dominates localization prospects, providing very tight uncertainty regions on the sky ($\sim10^{-5}$ deg$^2$ for SNR=10) due to rapid oscillations of the term $\sin[\omega L (1-\cos\xi)]$. In the latter, when pulsar distances are known imprecisely, antenna response variations provide the only recourse for GW source localization; these antenna response patterns are broad and slowly varying across the sky, such that localization prospects can be several orders of magnitude worse than when Earth-pulsar--term interference information is leveraged. We have derived the fully general analytic Fisher matrix determinant under this antenna-response--driven scenario, complementing earlier results in the small-angle limit between pulsar and GW source \citep{2012PhRvD..86l4028B}.

Our numerical studies have shown that for any practically-achievable pulsar distance precisions, the best localization prospects will occur when pulsars are in close proximity to (but not exactly atop) a GW source. As pulsar distance precisions are improved, GW source localization scales as $\Delta\Omega_\mathrm{sky}\propto\sigma_L^2$ until the Earth-pulsar system reaches its diffraction-limited saturation precision, which occurs at larger distance uncertainty values for pulsars in closer source proximity. All else being equal, distant pulsars contribute more to GW localization precision than nearby ones, such that $\Delta\Omega_\mathrm{sky}\propto 1 / L^2$.

We also studied an important variation of approaching the search for an individual GW signal in PTAs; whereas the \textit{phase-linked} approach treats the initial pulsar-term phase and frequency as derived entirely from binary and PTA properties, the \textit{phase-decoupled} approach abandons the phase connection, instead treating each initial pulsar-term phase as a nuisance variable over which we marginalize. This is how all current production-level PTA search pipelines function, and is a direct consequence of the cost-benefit ratio between navigating the highly oscillatory likelihood landscape under the phase-linked model, and the limited pulsar distance precisions we currently have. In effect, the latter makes the former not worth the bother. The result is localization behavior as a function of pulsar proximity that is similar to the antenna-response--driven scenario, since the only remaining impact of the pulsar term is through its frequency evolution; this is controlled by the factor $\dot\omega/\omega^2$, which is $\sim10^{-4}$ times smaller than potential pulsar-term phase contributions. Improved pulsar distance precisions have very little benefit to the precision of GW source localization in the phase-decoupled approach.

\begin{figure} 
    \includegraphics[width=\columnwidth]{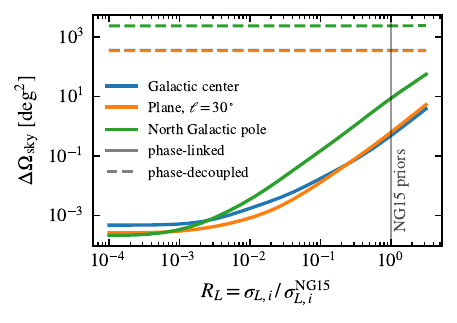}
    \caption{\label{fig:ng15_sigma_sweep}We take the positions, relative noise properties, distance values and precisions for the $67$ pulsars in the NANOGrav $15$-year dataset. These pulsar distance precisions are varied such that $\sigma_{L,i}=R_L\times\sigma^\mathrm{NG15}_{L,i}$. We examine the localization precision of GW sources placed in either the Galactic center, elsewhere in the plane at a Galactic longitude of $30^\circ$, or the North Galactic Pole, under the phase-linked \textit{(solid lines)} and phase-decoupled \textit{(dashed lines)} approaches.}
\end{figure}

To conclude, we examine the GW localization capabilities of a more realistic PTA with heterogeneous timing baselines and noise properties. We take $67$ pulsar positions, timing baselines, observational epochs, and residual rms values from the NANOGrav $15$-year dataset \citep[][hereafter NG15]{NG15_data}. We treat all noise as white, such that any intrinsic red or chromatic noise contributes only to the residual rms, which is converted to a weighted rms using the reported TOA uncertainties, then an effective rms that distributes the pulsar's noise among its observational epochs. We also take the real pulsar distance values and uncertainties, which vary in the ranges $0.156\leq (L_i / \mathrm{kpc})\leq 5.39$ and $10^{-3}\leq (\sigma_{L,i} / \mathrm{kpc})\leq 2$, respectively.

\begin{figure*} 
    \includegraphics[width=\textwidth]{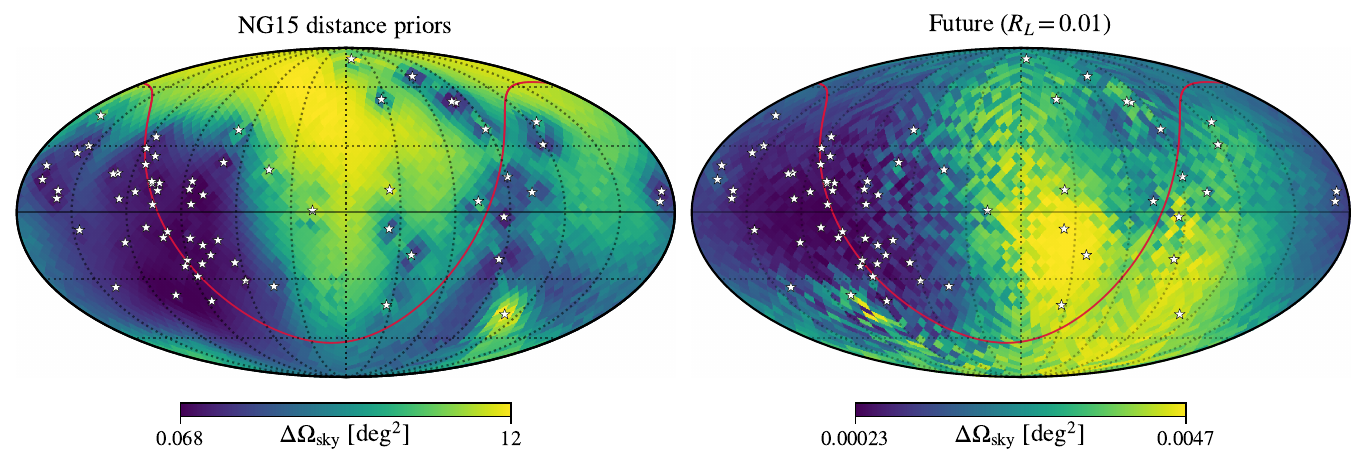}\\
    \includegraphics[width=\textwidth]{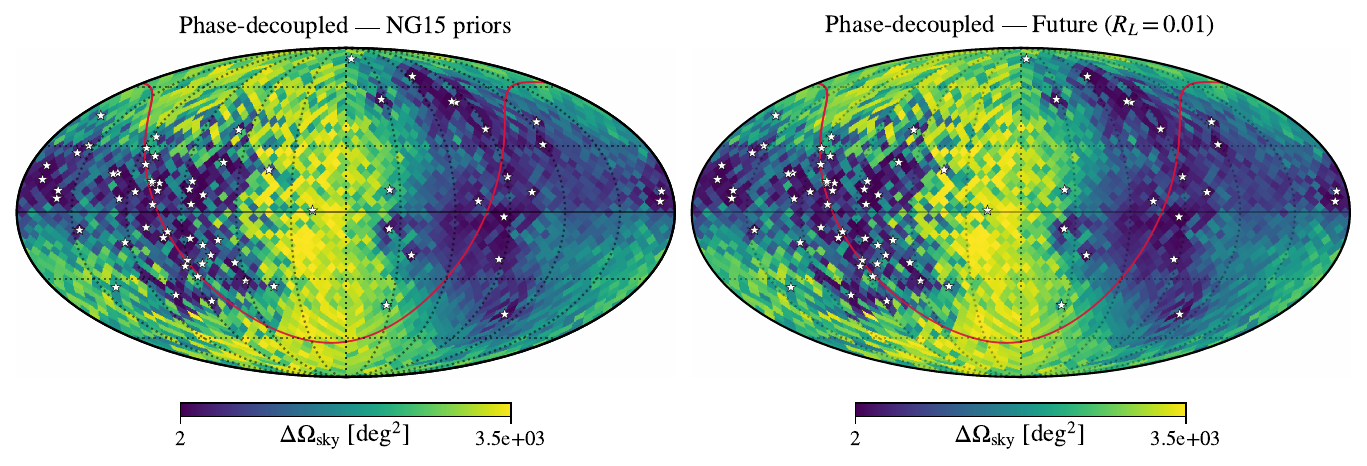}
    \caption{\label{fig:ng15_skymaps}Similar to \autoref{fig:ng15_sigma_sweep}, with $R_L=[0.01,1]$ for the phase-linked \textit{(top row)} and phase-decoupled \textit{(bottom row)} approaches. The Galactic plane is indicated by a red line in this coordinate system, with the $67$ NANOGrav $15$-year pulsars that were employed in GW searches shown as white stars.}
\end{figure*}

In \autoref{fig:ng15_sigma_sweep}, we apply the phase-linked and phase-decoupled approaches to GW sources in three different sky locations and for varied ratios, $R_L$, of pulsar distance uncertainties with respect to present-day NG15 values. The GW source properties are the same as elsewhere in this paper, and we maintain $\mathrm{SNR}=10$. The trends are similar to what we have seen in our idealized ring PTAs, with the GW localization performance of NG15 still mostly influenced by pulsar distance priors, and would only reach the array's diffraction-limited floor when $R_L\sim10^{-3}$. By contrast, the phase-decoupled performance shows no change with $R_L$. NG15's GW localization performance is shown more systematically as a function of source location in \autoref{fig:ng15_skymaps} for the phase-linked and phase-decoupled approaches, and with $R_L\in[0.01,1]$. Overall, the performance of NG15 with present-day production-level analysis pipelines (i.e., using the phase-decoupled approach) could achieve a best-case location precision of $\sim2$~deg$^2$, which would improve to $\sim 0.07$~deg$^2$ if the phase-linked approach were used. The NG15 diffraction-limited floor is capable of being $\lesssim 10^{-3}$~deg$^2$. Further improvements would require more pulsars to expand the array (which in itself seems more achievable than the pulsar distance uncertainty improvements necessary to reach the diffraction-limited floor).

There are important caveats to all of the studies presented here. We have been interested in GW source localization precision, not accuracy. Previous work has shown that accounting for pulsar terms in GW signal searches is critical for localization accuracy; neglecting them can lead to substantial biases. Moreover, even with the potential to localize GW source locations to within an interference fringe width, one is still left with the problem of identifying the correct fringe. We see this in our formulation of the $\overline{\mathrm{SNR}}$ detection statistic in \autoref{eq:snr}, which has many tightly packed interference fringes within a more broadly varying response envelope. Our Fisher formalism describes the \textit{local} curvature of the likelihood surface; in the non-evolving signal scenario with well-measured pulsar distances, demanding phase coherence can lead to very small localization areas, albeit with many repeated, neighboring solutions that have almost negligible penalty from the broader antenna response factor. This is partially mitigated by binary evolution over $\tau$, where one has to identify a sky location to match the pulsar-term phase \textit{and} the pulsar-term frequency; while a small effect, this can help to prune some of the repeated solutions encountered in the non-evolving phase-linked scenario \citep{2010arXiv1008.1782C}.

Since binary evolution over the baseline of PTA observations, $T$, is likely to be negligible, leveraging the pulsar term is the only way we can measure the signal chirp, and hence the binary's chirp mass for subsequent system characterization and host identification attempts. Improved pulsar distance precision can also be impactful on GW detection statistics, for which the studies of the present paper are a downstream consequence. Finally, one must have the usual caution when interpreting precision estimates derived under the Fisher formalism; it is possible that improvements to localization precision will be more pronounced for signals emerging from the detection threshold, and as parameter posteriors pull away from their prior distributions.

We close with a clarion call. In order for the conventional wisdom to be true that improved pulsar distance estimates will yield extraordinary PTA science, we must find a way to recouple the pulsar-term phase to binary and PTA characteristics in signal searches. This is exceptionally challenging, as the vast number of interference fringes in \autoref{fig:interference_viz} implies. Yet as prospects for resolving individual sources out of the confusion GW background improve, the benefits of surmounting this challenge call for a fresh inspection of the problem. This paper has served as a theoretical examination of GW source localization with PTAs and its various underpinning influences in aid of this next key milestone.

\begin{acknowledgments}
SRT thanks the anonymous referee for valuable suggestions that improved the quality of this work. SRT is grateful for discussions with Linqing Wen that inspired this work, and with Neil Cornish, who refined the history of phase-linked and phase-decoupled sampling approaches in Bayesian continuous GW searches. Jeff Hazboun, Pat Meyers, and Michael Lam commented on a preliminary draft, for which SRT is very grateful. SRT thanks Levi Schult and Polina Petrov, whose rigorous investigations provided motivation; William Lamb for cajoling the author into seeing the elegance of complex antenna response functions; Matt Miles for many fruitful discussions on interference between Earth terms and pulsar terms; and the entire VIPER group for their support. Elite Martial Arts of Bellevue is acknowledged for providing an outlet when certain derivations were eluding the author. 

SRT is a member of the NANOGrav collaboration, which receives support from NSF Physics Frontiers Center award numbers 1430284, 2020265, and 2607948. SRT acknowledges support from an NSF CAREER \#2146016, NSF AST-2307719, NSF NRT-2125764, and NASA LPS-80NSSC26K0342. SRT also acknowledges support from a Chancellor's Faculty Fellowship from Vanderbilt University.
\end{acknowledgments}

%

\software{\texttt{JAX} \citep{jax2018github}, \texttt{NumPy} \citep{harris2020array}, \texttt{SciPy} \citep{2020SciPy-NMeth}, \texttt{Matplotlib} \citep{Hunter:2007}, \texttt{HEALPix} \citep{2005ApJ...622..759G}, \texttt{healpy} \citep{Zonca2019}. Coding optimizations were made with Claude Opus/Fable through Github Copilot Pro in \texttt{VSCode}, and Google's \texttt{Antigravity} IDE.}


\bibliography{sample701}{}
\bibliographystyle{aasjournalv7}

\end{document}